\documentstyle[12pt,epsfig]{article}
\begin{document}

\title {\bf Coordinated Slowing of Metabolism in Enteric Bacteria
under Nitrogen Limitation: A Perspective}

\author{Ned S. Wingreen\\
NEC Research Institute, 4 Independence Way\\
Princeton, New Jersey 08540\\
and Department of Physics, University of California\\ 
Berkeley, CA 94720\\
\ \\
Sydney Kustu\\
Department of Plant Biology, Molecular 
and Cell Biology\\ 
University of California, Berkeley, CA 94720}

\date{}

\maketitle

\begin{abstract}

It is natural to ask how bacteria coordinate metabolism when depletion of
an essential nutrient limits their growth, and they must slow their entire
rate of biosynthesis.  A major nutrient with a
fluctuating abundance is nitrogen. The growth rate of enteric bacteria
under nitrogen-limiting conditions is known to correlate 
with the internal concentration of free glutamine, the glutamine pool.
Here we compare the patterns of utilization of L-glutamine and L-glutamate, 
the two central intermediates of nitrogen metabolism. Monomeric precursors of
all of the cell's macromolecules -- proteins, nucleic acids, and surface
polymers -- require the amide group of glutamine 
at the first dedicated step of biosynthesis.
This is the case even though only a minority ($\sim$12\%) 
of total cell nitrogen 
derives from glutamine.  In contrast, the amino group of glutamate,
which provides the remainder of cell nitrogen, is
generally required late in biosynthetic pathways, {\it e.g.} in 
transaminase reactions for amino acid synthesis.
We propose that the pattern of glutamine dependence coordinates the decrease
in biosynthesis under conditions of nitrogen limitation.    
Hence, the glutamine pool plays a global regulatory role in the cell.
\end{abstract}

\newpage

\baselineskip=0.8cm

\noindent INTRODUCTION
\bigskip

Enteric bacteria are notable for their varying environment.
Within a host they can experience conditions of high nutrient
availability, high osmolarity, and high toxicity; outside a host, 
they can encounter the opposite. Evolution has adapted enteric 
bacteria to grow and divide, albeit at different rates, under both
conditions. One can reasonably ask ``How does the metabolism
of these bacteria remain in balance under such different conditions?" 
Here we consider this question in relation to the response of 
{\it E. coli} and {\it S. typhimurium} to nitrogen availability.

We chose to consider nitrogen metabolism for several reasons.
First, central metabolism of
nitrogen is well studied and is relatively simple compared to carbon
metabolism:  there are only two central intermediates in nitrogen
metabolism, the amino acids glutamate and glutamine, and these are
synthesized from a single precursor, 2-oxoglutarate.  In contrast, 
carbon metabolism entails the synthesis of 12 central intermediates. 
Second, all assimilated nitrogen is used for biosynthesis, so it is not 
necessary to consider storage forms or waste products. 
Finally, enteric bacteria have a strong nitrogen starvation response.  
It appears that transcription of up to 2\% of the 
{\it E. coli} genome is activated by nitrogen regulatory protein C 
(NtrC) \cite{Zimmer00}.

\begin{figure}[t]
\centering
\epsfig{file=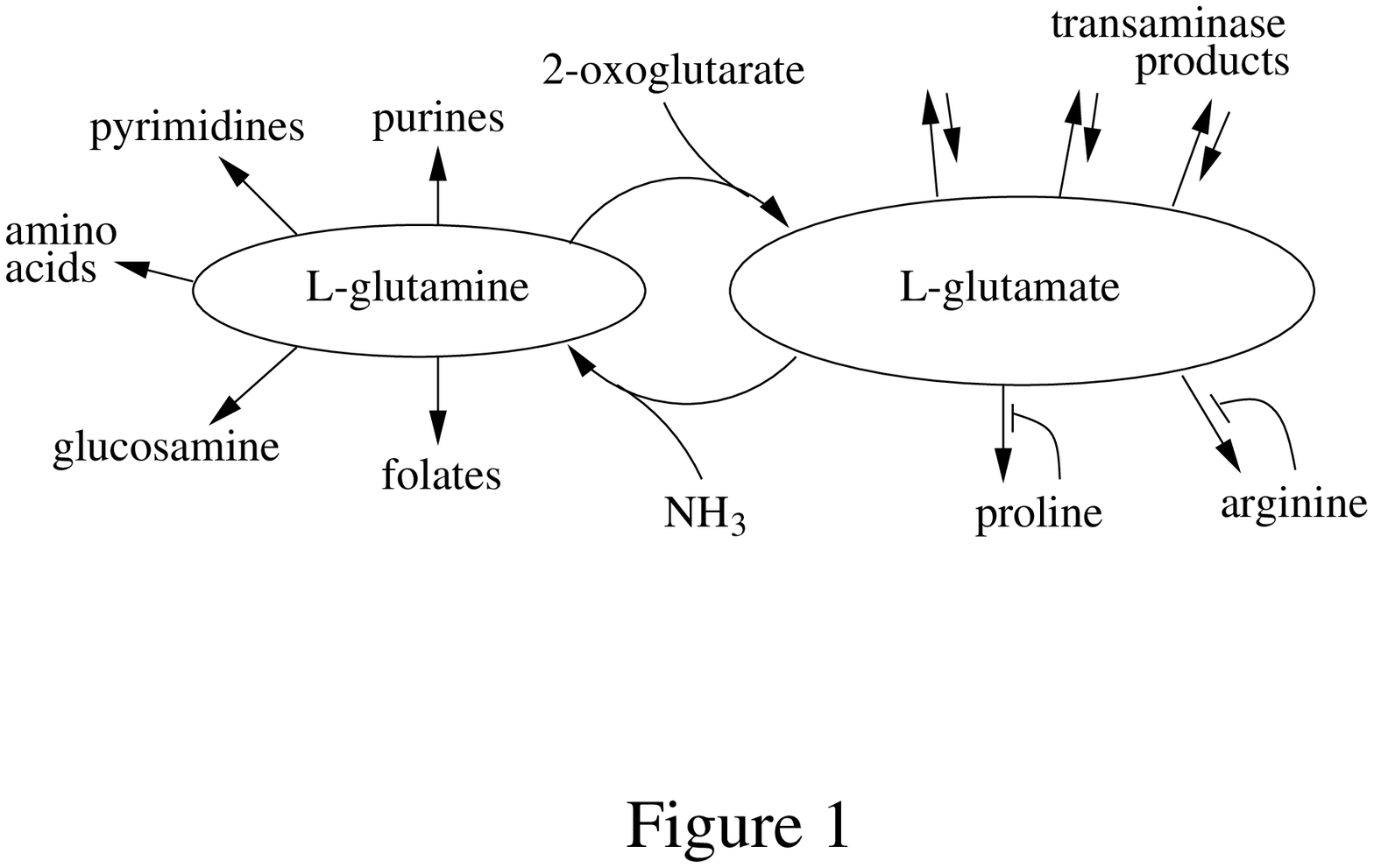,width=1.0\linewidth}
\caption{ 
Nitrogen central intermediates L-glutamine and L-glutamate. Under 
nitrogen-limited conditions, nitrogen is incorporated via the 
GS/GOGAT cycle as shown. The metabolism of glutamine is the 
source of only $\sim12\%$ of the nitrogen content of the cell. 
Nevertheless, L-glutamine is required for the first dedicated
step in the biosynthesis of monomeric precursors of all the
cell's macromolecules - purines and pyrimidines for nucleic
acids, amino acids for proteins, and glucosamine for surface
polymers.  In contrast, L-glutamate is generally required late
in biosynthetic pathways. For a more complete listing of 
reactions requiring L-glutamine or L-glutamate, see Tables 1 and
2, respectively.}
\label{fig:GS}
\end{figure}

Enteric bacteria assimilate inorganic NH$_3$ into organic molecules
by two means. In the reaction catalyzed by biosynthetic glutamate
dehydrogenase, NH$_3$ is assimilated directly into glutamate.
In the reactions of the glutamine synthase/glutamate synthase
(GS/GOGAT) cycle, which is widely present in prokaryotes, NH$_3$ is
assimilated first into glutamine and then into glutamate 
(Fig.~\ref{fig:GS}).
Under nitrogen-limited growth conditions, 
the major assimilatory pathway is the glutamine 
synthase/glutamate synthase (GS/GOGAT) cycle \cite{Tempest70,Reitzer90}.
If one traces biosynthetic nitrogen back to its last appearance 
in one of the two central intermediates, 
then only $\sim 12\%$ of cellular nitrogen derives from glutamine. 
The remainder comes from glutamate \cite{Wohlhueter73}.  
Nevertheless, enteric bacteria appear to perceive external nitrogen 
limitation as internal {\it glutamine} limitation \cite{Ikeda96}. 
For a variety of nitrogen-limiting conditions, the growth rate of 
enterobacteria was found to correlate with the glutamine 
pool, which dropped by a factor of 10 or more. The glutamate pool
remained high.

Motivated by the above observations, we explored whether the metabolic 
uses of glutamine might be better suited than those of 
glutamate to slowing biosynthesis in a coordinated manner. It is
remarkable that glutamine is required at the first dedicated step 
in the biosynthesis of many of the cell's major biosynthetic monomers:  
amino acids, purines and pyrimidines, and UDP-GlcNAc, a 
precursor of the cell-surface polymers murein and 
lipopolysaccharide. 
We consider the consequences of this pattern of glutamine utilization 
and propose that the uses of glutamine in metabolism, coupled
to the drop in the glutamine pool, may be important in maintaining
homeostasis under nitrogen-limiting conditions. 

\bigskip
\noindent METHODS    
\bigskip

To compare patterns of glutamine and glutamate 
utilization in metabolism, we made use of the following
sources:  
{\it Escherichia coli and Salmonella typhimurium}
edited by Neidhardt \cite{Neidhardt96},
and references contained therein; 
{\it The Enzymes of Glutamine Metabolism} edited by  
Prusiner and Stadtman \cite{Prusiner73};
and {\it The Amidotransferases} by Zalkin \cite{Zalkin93}.
We made extensive use of original literature, particularly
for $K_m$ values for the enzymes of glutamine and glutamate 
metabolism (Tables 1 and 2). Because these apparent $K_m$
values -- concentrations required for half-maximal velocity --
were measured {\it in vitro} under a variety of conditions,
they can be regarded as only approximate guides to {\it in vivo} 
properties of the enzymes. 
Two easily accessible Web databases and portals that we utilized are 
EcoCyc, a database of metabolic pathways in {\it E. coli} 
(http://ecocyc.pangeasystems.com/ecocyc/),
and KEGG, a database of metabolic pathways for a collection
of organisms\\
(http://www.genome.ad.jp/kegg/).

\bigskip
\noindent RESULTS    
\bigskip

Glutamine and glutamate play different roles in biosynthesis.
(i) Though the source of only 12\% of cell nitrogen, the amide
group of glutamine is required directly for the biosynthesis 
of monomeric units of all macromolecules (proteins, 
nucleic acids, and surface polymers).
(ii) With the exception of the histidine biosynthetic pathway, 
glutamine is required at the first dedicated step of every 
pathway in which it is utilized. By contrast, glutamate is
required more often than not at the end or in the middle 
of pathways. (iii) Glutamine-dependent reactions are generally
coupled to ATP hydrolysis, and hence are essentially irreversible. 
More than half of the reactions involving glutamate are
reversible transaminase reactions.

We first consider the utilization of glutamine.
The enzymes of glutamine metabolism are listed in
Table 1, along with their $K_m$ values for glutamine.
Co-reactants for these enzymes and the conventional
pathway designations are also listed. 
The $K_m$ values for glutamine of the enzymes
fall in the range 0.2 - 1.7 mM. 
The glutamine pool concentration of {\it S. typhimurium}
dropped from approximately 3-4 mM on ammonia, the optimal
nitrogen source, to $< 0.3$ mM on growth-rate-limiting 
nitrogen sources, or in continuous culture with ammonia
as the limiting nutrient. Comparison of pool sizes to 
$K_m$ values provides evidence that the velocities of
many enzymes that utilize glutamine fall below 
half-maximal under nitrogen-limiting conditions.
The fluxes in the corresponding biosynthetic pathways will
be concomitantly reduced.

\begin{figure}[t]
\centering
\epsfig{file=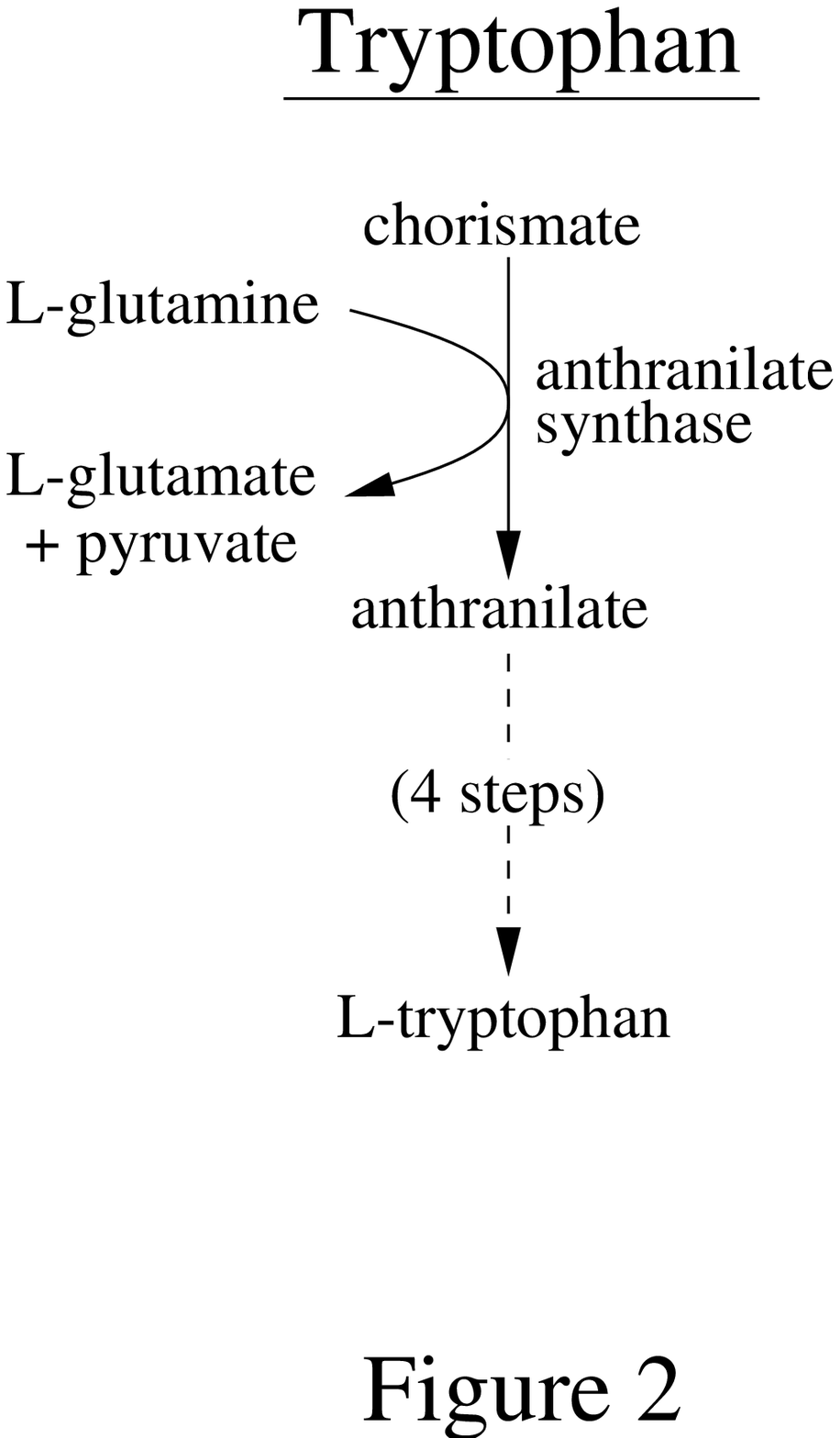,width=0.4\linewidth}
\caption{
Tryptophan biosynthetic pathway. The first dedicated step of 
tryptophan biosynthesis requires glutamine.}
\label{fig:trp}
\end{figure}

\begin{figure}[t]
\centering
\epsfig{file=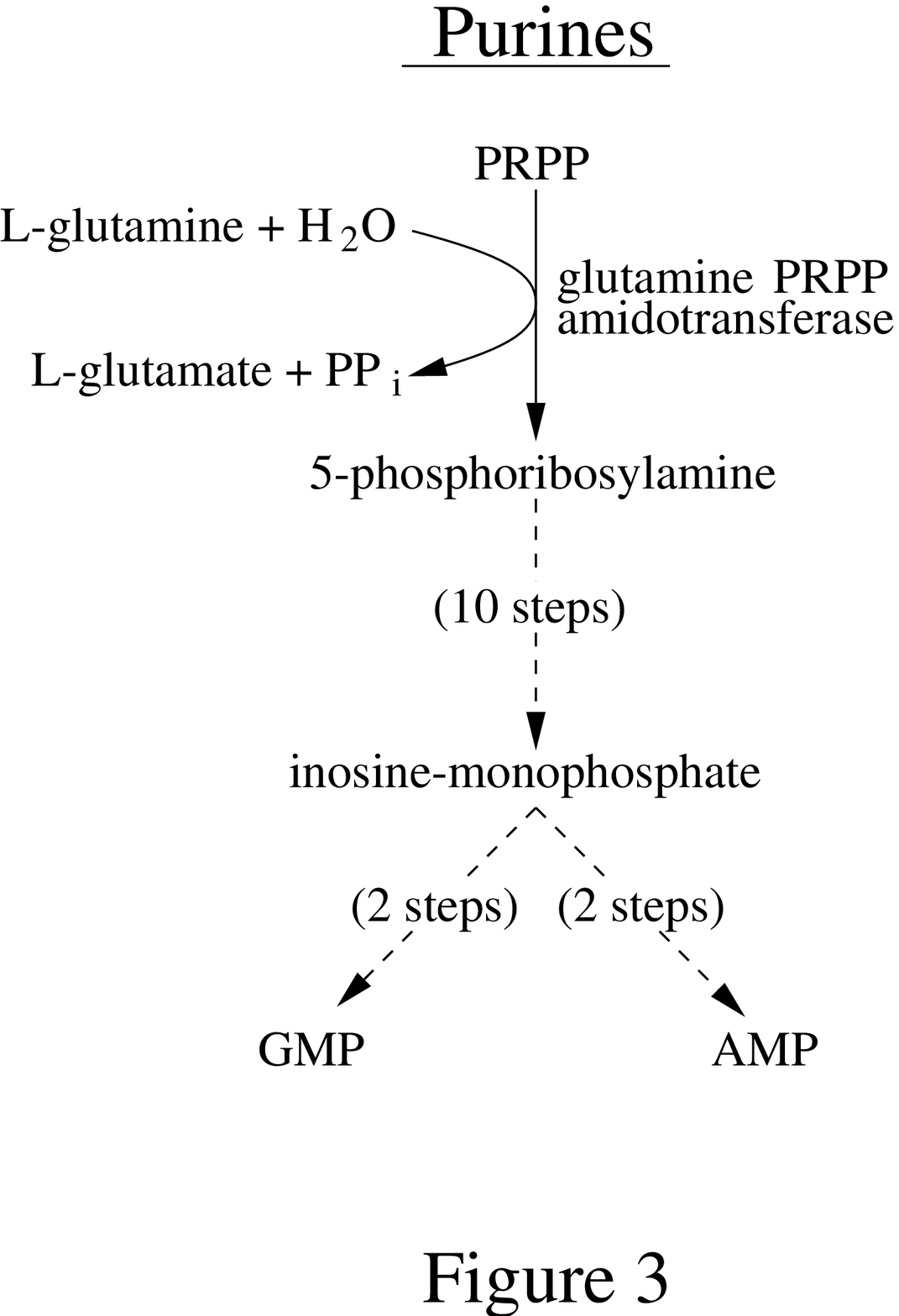,width=0.6\linewidth}
\caption{
Purine biosynthetic pathway. The first dedicated step of 
purine biosynthesis requires glutamine. The fourth step 
also requires glutamine, as does the final step
in GMP biosynthesis.}
\label{fig:pur}
\end{figure}

\begin{figure}[t]
\centering
\epsfig{file=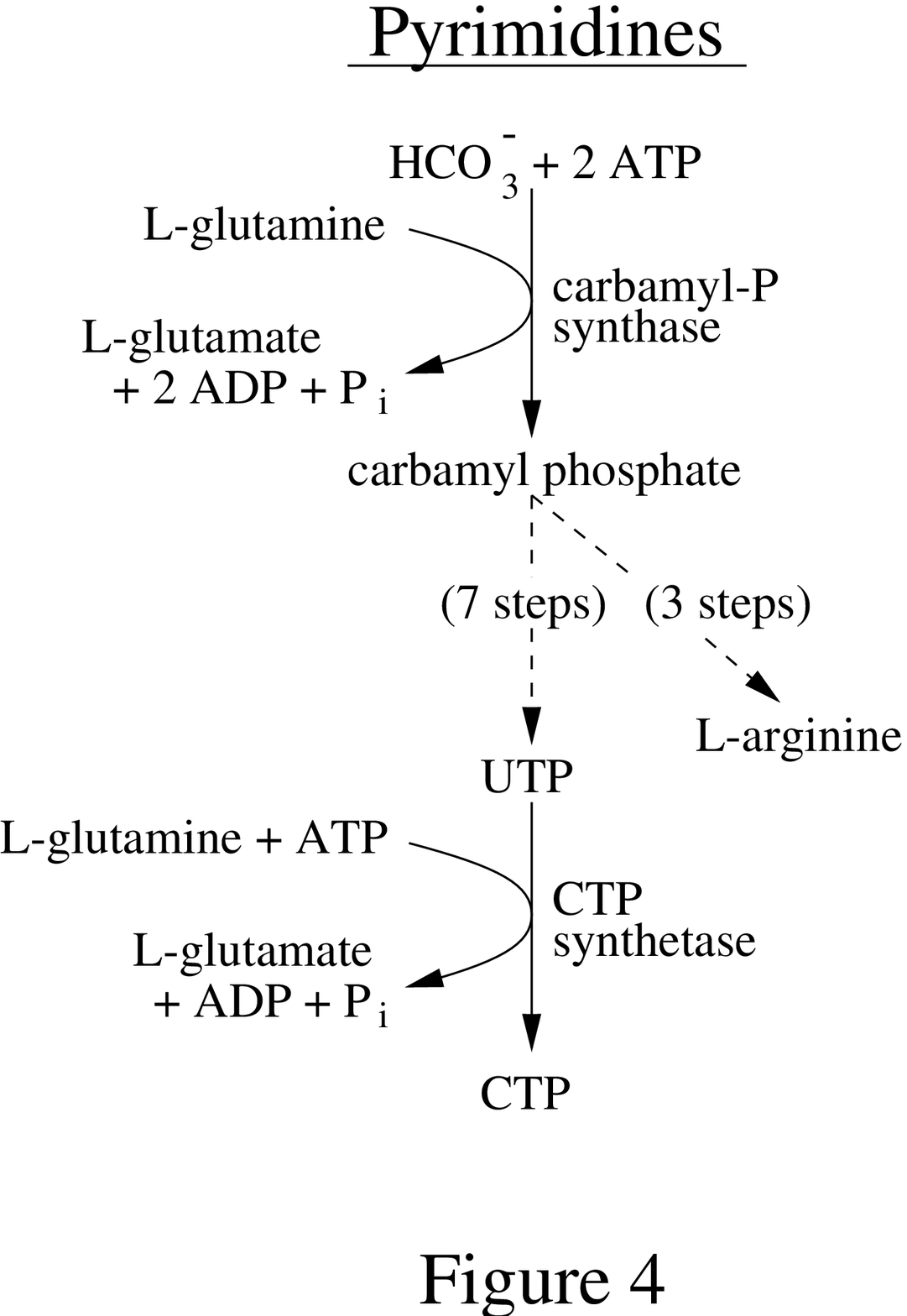,width=0.6\linewidth}
\caption{
Pyrimidine biosynthetic pathway. Glutamine is required for 
the biosynthesis of carbamyl phosphate, which is a parent 
compound for both pyrimidine and arginine biosynthesis. The final
step in CTP biosynthesis also requires glutamine.}
\label{fig:pyr}
\end{figure}

\begin{figure}[t]
\centering
\epsfig{file=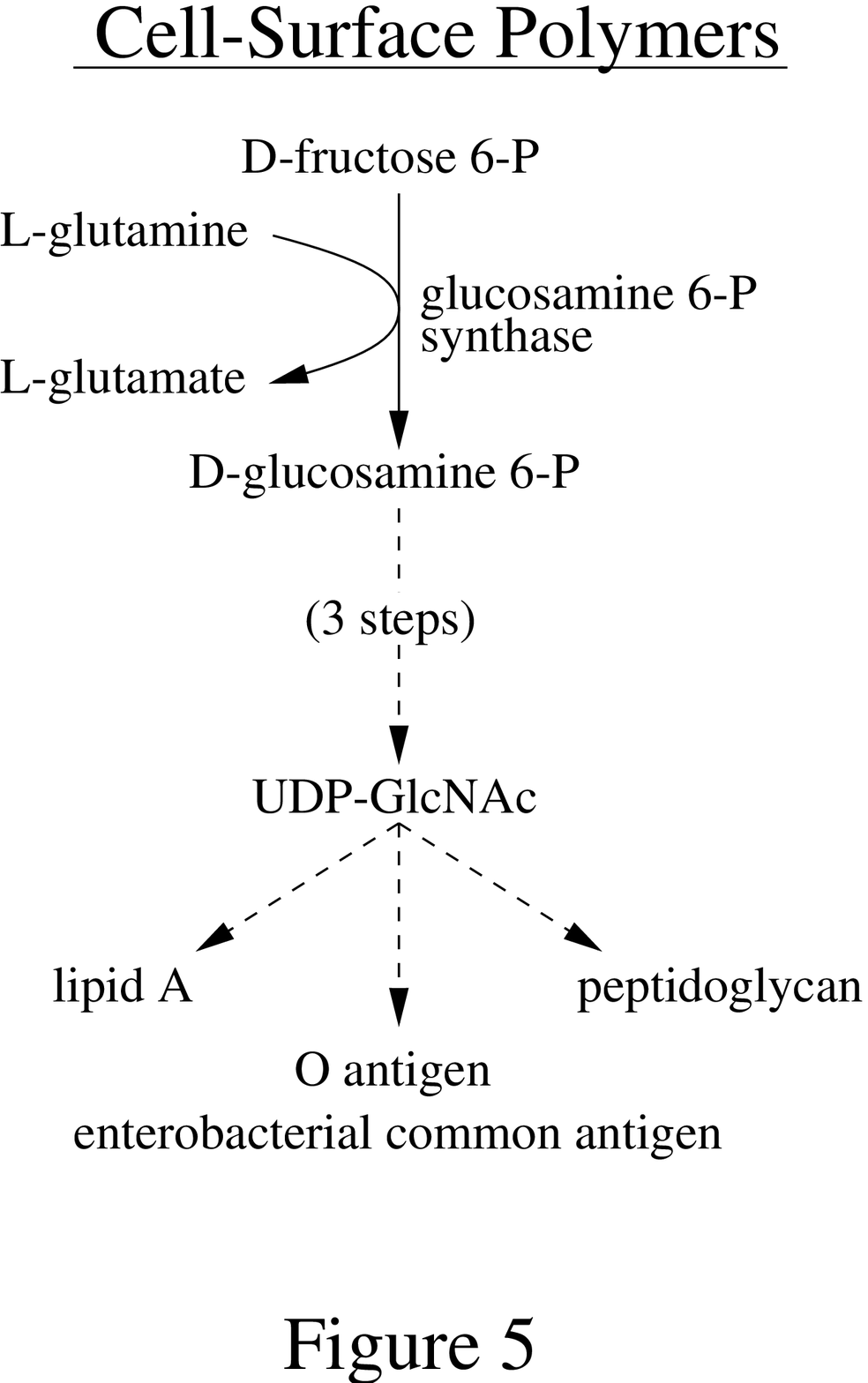,width=0.5\linewidth}
\caption{
Cell-surface polymer biosynthetic pathways. Glutamine is 
required for the first dedicated step of the biosyntheses
of peptidoglycan (murein), lipid A, O antigen, and 
enterobacterial common antigen.}
\label{fig:pol}
\end{figure}

\begin{figure}[t]
\centering
\epsfig{file=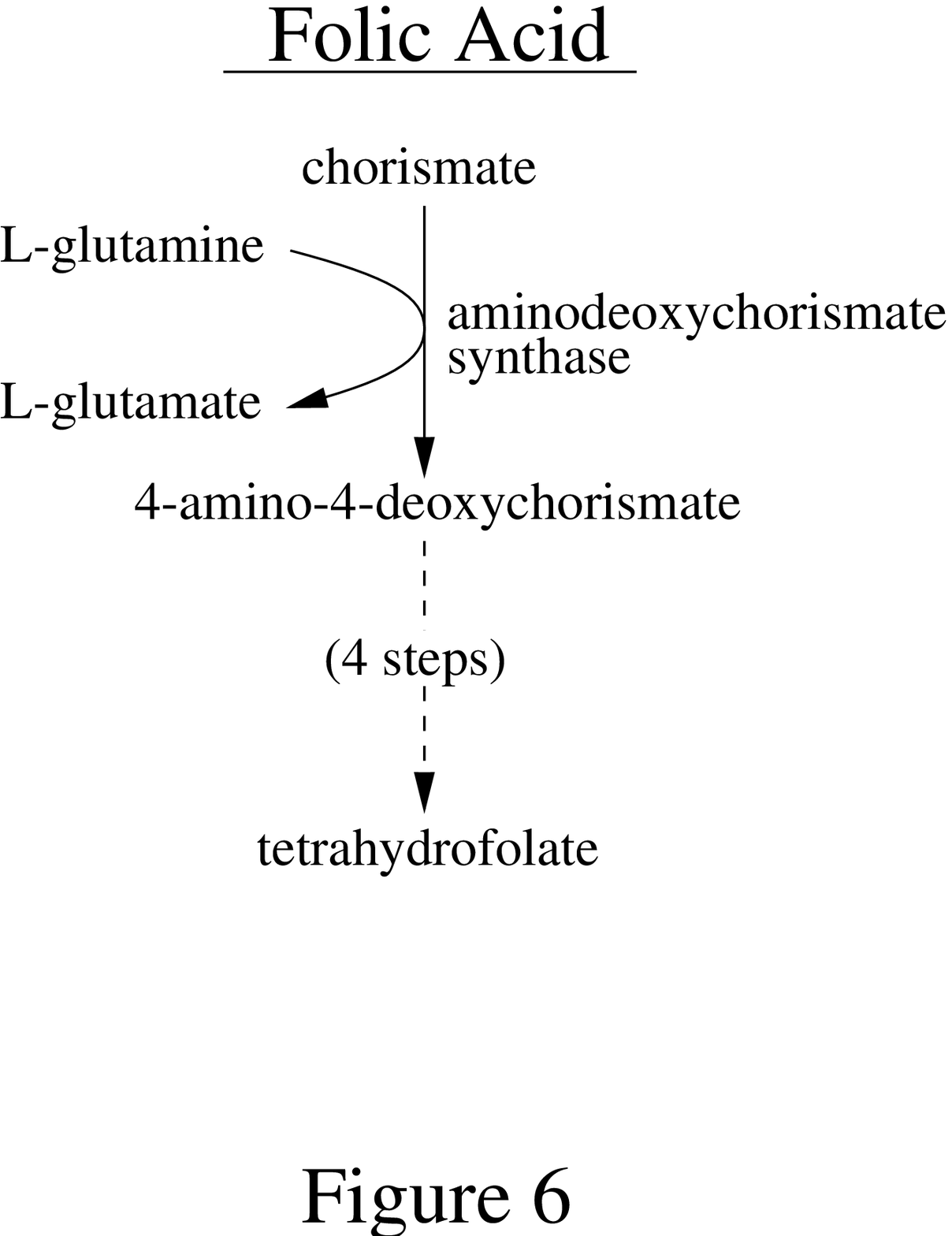,width=0.5\linewidth}
\caption{
Folic-acid biosynthesis. The first dedicated step of folic-acid 
biosynthesis requires glutamine.}
\label{fig:fol}
\end{figure}

\begin{figure}[t]
\centering
\epsfig{file=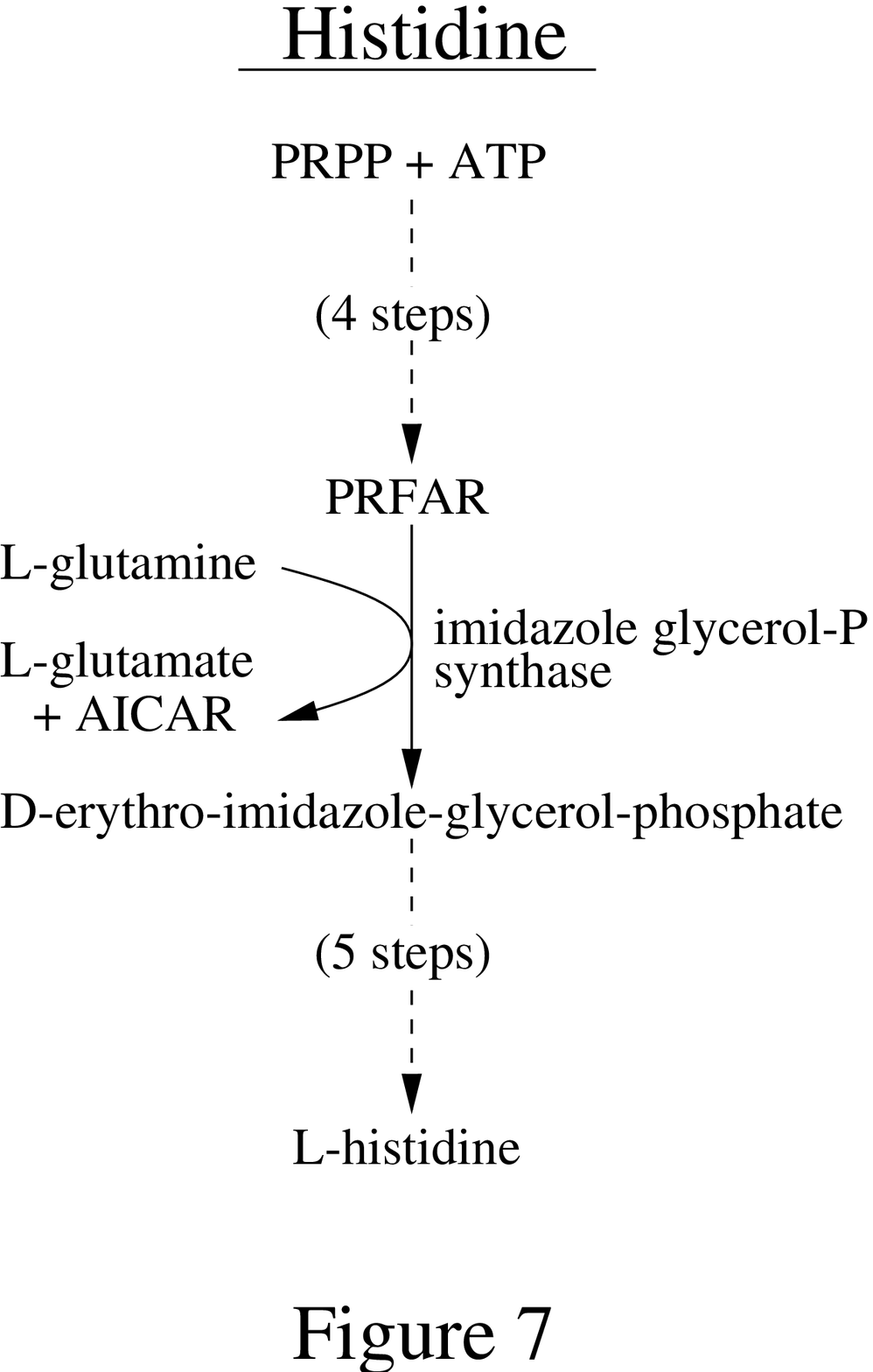,width=0.5\linewidth}
\caption{
Histidine biosynthesis. Glutamine is required for the fifth 
dedicated step of histidine biosynthesis.}
\label{fig:his}
\end{figure}

Figures~\ref{fig:trp}-\ref{fig:his} diagram the biosynthetic pathways
that require glutamine. Glutamine is required directly
for the biosynthesis of all of the major macromolecules
of the cell: proteins (via the amino acids tryptophan,
histidine, arginine, and glutamine itself),
nucleic acids (via purines and pyrimidines), 
and cell-surface polymers (via glucosamine). 
Glutamine is also required for the biosynthesis of the 
cofactor folic acid.  In all of these cases,
glutamine acts as a ``nitrogen carrier", losing its amide
group and hence yielding glutamate as an immediate product.

In all but one case, the glutamine-dependent reaction is the 
first dedicated step in a biosynthetic pathway or the first 
dedicated step in synthesis of a parent compound of the pathway 
({\it e.g.} carbamyl phosphate for the arginine and pyrimidine 
pathways). For example, as shown in Fig.~\ref{fig:trp}, glutamine 
is a substrate for the first reaction unique to tryptophan 
biosynthesis.  The other substrate, chorismate, is a precursor 
for many other aromatic compounds. Likewise, glutamine is
required at the first dedicated step in purine biosynthesis. For this 
reaction, the other substrate, phosphoribosyl pyrophosphate (PRPP),
is a common precursor for several pathways, including the
histidine, tryptophan, and pyrimidine biosynthetic pathways.
The fourth dedicated step of purine biosynthesis also requires
glutamine, as does the biosynthesis of guanosine monophosphate 
(GMP) from xanthosine monophosphate (XMP). 

Figure~\ref{fig:pyr} shows the glutamine-dependent steps in the pyrimidine
pathway. Glutamine is required for 
the biosynthesis of carbamyl phosphate. 
The latter is a parent compound for 
both the pyrimidines and for arginine. Glutamine
is also required in the pyrimidine pathway 
at the final step from UTP to CTP.

The major intermediate in the biosynthesis of surface 
polymers is the compound UDP-GlcNAc. 
Initiation of the synthesis of peptidoglycan (murein), lipid A, 
O antigen, and enterobacterial common antigen all 
require UDP-GlcNAc \cite{Raetz96}. 
As shown in Fig.~\ref{fig:pol},  the synthesis of
UDP-GlcNAc requires the amino sugar 
D-glucosamine 6-phosphate, which is a product of
glutamine and the central metabolite
D-fructose 6-phosphate \cite{Zalkin93}.
Hence, glutamine is required at the first dedicated
step in the biosynthesis of the cell-surface polymers.

Glutamine is also required for the biosynthesis
of the folic-acid family of coenzymes. As shown
in Fig.~\ref{fig:fol}, glutamine enters at the first dedicated
step of the folic-acid pathway. The other substrate is
the common precursor chorismate. 

The utilization of glutamine in histidine
biosynthesis breaks with the above pattern. As shown in
Fig.~\ref{fig:his}, glutamine is required at the {\it fifth} step
of the dedicated histidine pathway. The enzyme
which catalyzes this reaction has a $K_m$ for glutamine 
of 0.24 mM, the lowest of the enzymes of glutamine
metabolism. This is below the concentration of
glutamine measured by Ikeda {\it et al.} Therefore,
the velocity of this reaction may not depend strongly
on glutamine in the range of nitrogen limitation studied. 

Finally, the amino acid asparagine is synthesized directly from 
glutamine, aspartate, and ATP. There is also an enzyme 
aspartate-ammonia ligase in E. coli that utilizes NH$_3$  
in place of glutamine to synthesize asparagine.
However, strains lacking the glutamine-dependent enzyme
asnB cannot grow in nitrogen-limited media \cite{Reitzer96}.

We consider now the utilization of glutamate. The enzymes
of glutamate metabolism are listed in Table 2. They are
divided into two classes: transaminase reactions and all 
others. The co-reactants and pathways are listed, along with 
the $K_m$ values for glutamate of the enzymes where available.
Most of the measured $K_m$ values for glutamate are 
roughly 10 times as large as those for glutamine. This is consistent
with the observation that the glutamate pool stays fixed in
the range 15-25 mM, approximately 6 times larger than the  
glutamine pool of 3-4 mM under unlimited nitrogen \cite{Ikeda96}.

In all the transaminase reactions, glutamate loses its amine
group yielding 2-oxoglutarate as one product. For these
reactions, glutamate plays the role of a nitrogen carrier,
similar to the role played by glutamine.  In contrast, the
entire glutamate molecule is assimilated in the non-transaminase
reactions. 

\begin{figure}[t]
\centering
\epsfig{file=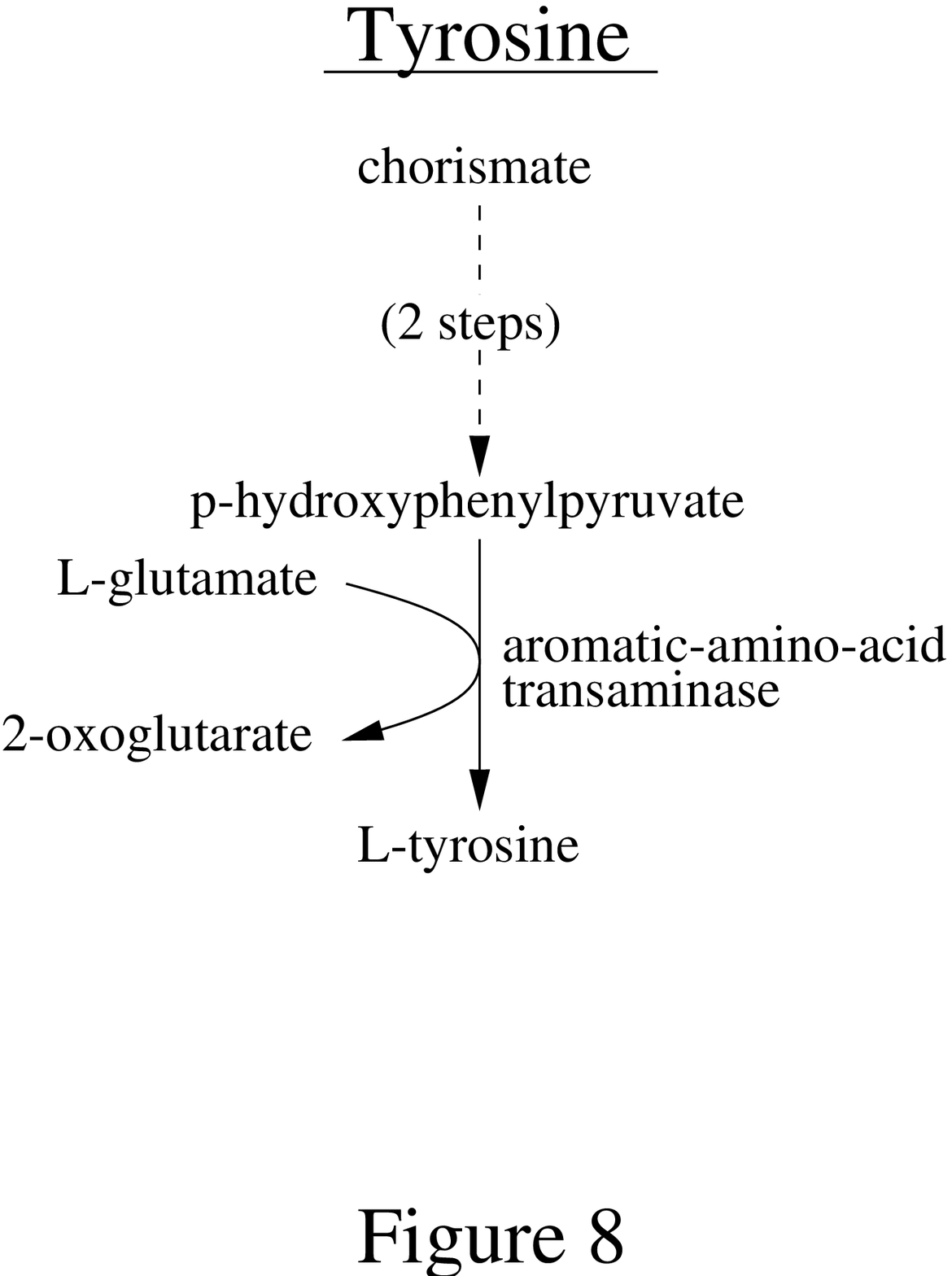,width=0.6\linewidth}
\caption{
Tyrosine biosynthesis. Glutamate is required at the last
step in the pathway for a transaminase reaction.}
\label{fig:tyr}
\end{figure}

In the interests of space, we have chosen to show only a
few characteristic examples of metabolic pathways involving 
glutamate. Figure~\ref{fig:tyr} shows the tyrosine pathway,
starting from the common precursor chorismate. 
The reaction involving glutamate is a reversible transaminase 
reaction and occurs at the final step of the pathway. As shown in
Table 2, there are twelve other pathways in which glutamate 
is involved via a transaminase reaction. In nine of the
twelve, the glutamate dependent step is later than the first 
dedicated step of the pathway. The cases in which 
glutamate is required for the first dedicated step are the
aspartate pathway and the valine pathways, both of which
are one-step pathways from common precursors,
and the enterobacterial common antigen (ECA) pathway. 
In the ECA pathway, the co-substrate with glutamate, 
dTDP-4-dehydro-6-deoxy-D-glucose, is also an intermediate
in the biosynthesis of rhamnose, a common constituent
of gram-negative O-antigens \cite{Stern99}.

\begin{figure}[t]
\centering
\epsfig{file=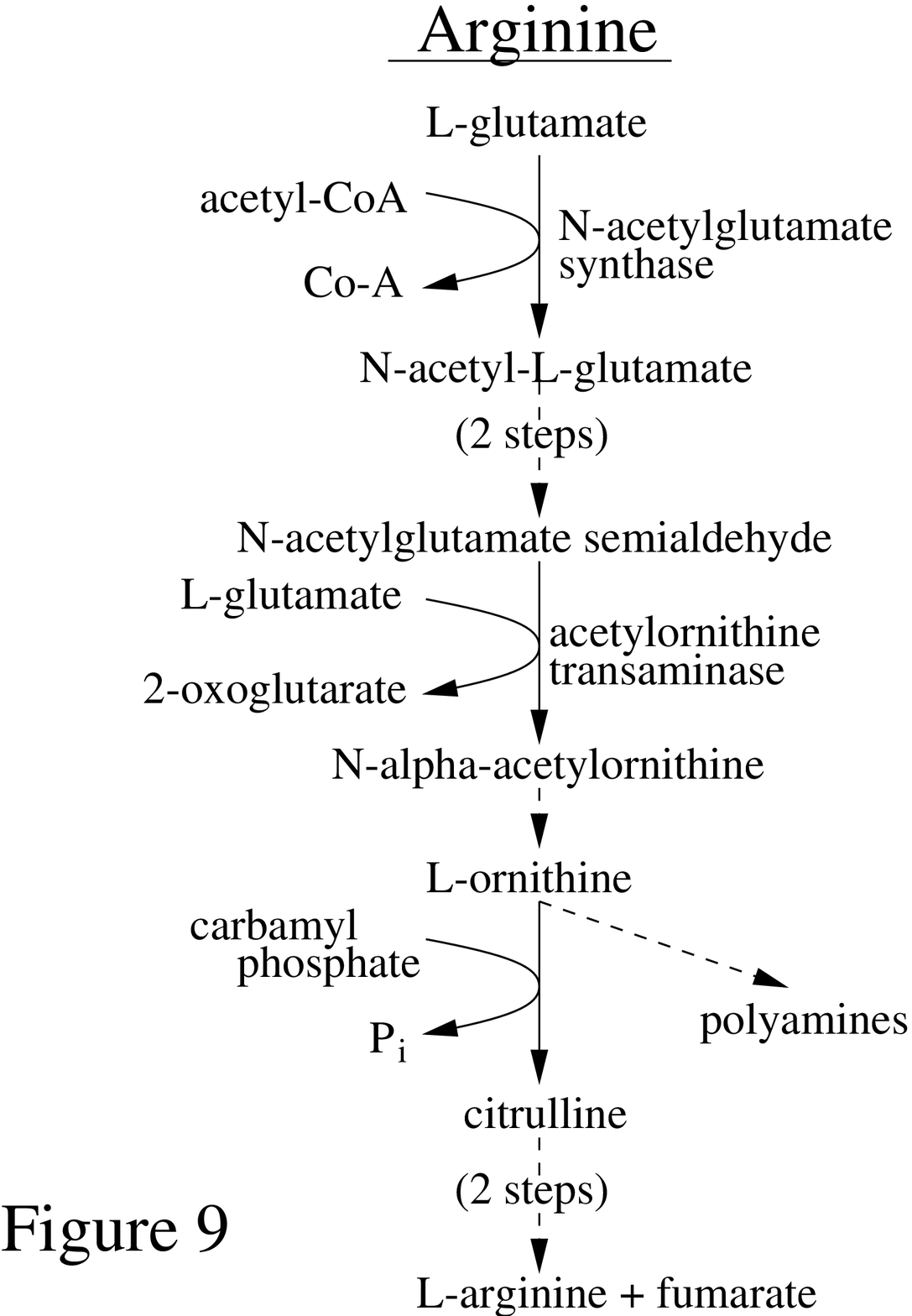,width=0.6\linewidth}
\caption{
Arginine biosynthetic pathway. Glutamate is the precursor
and is also required at the fourth step for a transaminase reaction.
Ornithine is a precursor for both arginine and the polyamines
putrescine and spermidine. The co-reactant with ornithine, 
carbamyl phosphate, is a direct product of glutamine (see Fig.~4)}
\label{fig:arg}
\end{figure}

Figure~\ref{fig:arg} shows the arginine pathway. Glutamate is the precursor 
compound and is also required for a transaminase reaction at the 
fourth step. Ornithine, an intermediate product in the arginine
pathway, is also a precursor for the polyamines putrescine and
spermidine \cite{Davis92}.

As shown in Table 2, glutamate is required in seven non-transaminase
reactions. Of these, the two in the folic-acid pathway occur later than 
the first dedicated step of the pathway. 
In contrast to the transaminase reactions, five of the non-transaminase 
reactions are coupled to ATP hydrolysis, 
and hence are expected to be irreversible in the cell. 

Essentially all nitrogen in cell products derives ultimately from glutamine 
or glutamate. For example, amino groups of amino acids not
listed in Table~2 still come indirectly from glutamate.
Specifically, aspartate and serine 
provide nitrogens in the synthesis of several other
amino acids, and their nitrogens are derived from glutamate. 
Unlike the reversible transaminase
reactions involving glutamate, the amino-acid
biosynthetic reactions involving
aspartate or serine are either coupled to hydrolysis of ATP,
or involve incorporation of the entire amino-acid 
molecule into the final product.

\bigskip
\noindent DISCUSSION
\bigskip

Several aspects of glutamine utilization appear to suit it to a general
metabolic regulatory role.  
While the amide group of glutamine donates only 12\% of cellular
nitrogen (with the remainder derived from glutamate),
glutamine dependent monomers are required to initiate
the syntheses of the cell's macromolecules. 
Primary depletion of the free pool
of glutamine under nitrogen-limiting conditions would therefore 
presumably slow fluxes into a number of biosynthetic pathways
simultaneously. As a consequence of glutamine dependence at
the first dedicated step in these pathways, no useless and/or toxic 
intermediates would accumulate. Instead, 
any accumulation would be of precursors for the
pathways, {\it i.e.} substrates for their first enzymatic steps.  These are
substances like chorismate, PRPP, D-fructose 6-P, and HCO$_3^-$, which are
common precursors for several pathways and are readily converted into
other central metabolites ({\it e.g.} ribose 5-P and ATP from PRPP). In
addition, many of the glutamine-utilizing reactions are essentially
irreversible because they are coupled to hydrolysis of ATP. Thus,
these reactions cannot reverse to refill a glutamine pool
depleted by a decrease in ammonia assimilation.  Glutamine
product pools can therefore be maintained high -- thus preventing
product starvation responses -- even in the presence of a depleted 
glutamine pool. Finally, maintenance of a high
glutamate pool under nitrogen-limiting conditions would insure that
synthesis of biosynthetic monomers, once initiated, would be completed.
There are transaminase-type reactions 
in both the purine and arginine pathways that utilize the amino 
group of aspartate rather than that of glutamate. All these reactions
are coupled to hydrolysis of ATP or GTP, thus also assuring completion 
of monomer synthesis.

By contrast, if depletion of the {\it glutamate} pool were the primary
consequence of decreased ammonia assimilation, the result would
be the accumulation of useless and/or toxic
intermediates in many biosynthetic pathways, {\it e.g.} 
folic-acid, histidine, iso-leucine, and lysine pathways. The presence of
excess glutamine under these conditions would exacerbate the
accumulation of such intermediates in pathways initiating
with glutamine, {\it e.g.} the intermediate 7,8-dihydropteroate
and its precursors in the folic-acid pathway. 
Moreover, the reversibility of the transaminase reactions 
utilizing glutamate would 
tend to replenish a depleted glutamate pool and would carry down
the pools of its products. A drop in product pools would induce 
product starvation responses, including upregulation of
enzymes metabolizing glutamate, resulting in a futile tug-of-war
with overexpressed enzymes competing for the depleted glutamate
pool.

If the free pool of glutamine does indeed regulate
biosynthetic fluxes, 
we believe it is critical to determine as 
directly as possible which pool or pools
of glutamine-derived products are, like the pool of glutamine itself,
actually depleted under nitrogen-limiting conditions.  
It is these secondary depleted pools which will directly limit 
the growth of the cell.
Knowledge of which glutamine-product pools are depleted will give
insight into the mechanisms the cell has adopted for slowing growth
without disrupting, for example, the expression of proteins
responsible for nitrogen assimilation.
(The $K_m$ value for glutamine of glutaminyl-tRNA synthetase
is measured to be 0.15-0.21 mM \cite{Kern80,Ibba96}, 
well below the glutamine concentrations measured 
for slow-growing cells \cite{Ikeda96}, 
so direct glutamine-dependent slowing of protein synthesis 
can be discounted as a primary cause of slow growth.) 

At least one glutamine-product pool must become depleted 
under nitrogen limitation or there would be no slowing of
growth. However, it is unclear whether glutamine depletion 
propagates into depletion of multiple product pools, 
or into depletion of just a single product pool. 
The depletion of multiple product pools would provide 
a simple functional explanation for why so many 
biosynthetic pathways depend on glutamine. 
Moreover, a slowing of growth by depletion
of multiple pools would likely be robust to fluctuations in 
product pool sizes. 

If only a single glutamine-product pool is depleted, the slowing of
growth would result in reduced demand for other cell products and,
potentially, large and measurable increases in other product pools. In
this scenario, the widespread dependence of biosynthesis on glutamine
would function to reduce these accumulations by directly reducing
biosynthetic rates.  Insofar as growth rate is limited by the overall
rate of nitrogen assimilation, no obvious growth advantage would
accrue from a mutation to increase the pool of the one depleted
product.

{\it Nucleotides and Nucleic Acids --}
Glutamine is a substrate for the first enzymatic steps in
synthesis of ATP, GTP, UTP, and CTP.  A significant 
drop in the pool concentrations
of these nucleotides would presumably decrease rates of
synthesis of both RNA and DNA.  A global decrease
in $m$RNA synthesis might, in turn, decrease protein synthesis. 
However, genes of the nitrogen-starvation response are known to be
upregulated under nitrogen limitation \cite{Zimmer00},
and would have to escape such a global decrease in $m$RNA
synthesis. ATP and GTP are widely employed by the cell
as energy carriers, so a significant drop in their pools would 
have a widespread effect on metabolic rates. CTP, which is
a direct glutamine-product from UTP (Fig.~\ref{fig:pyr}), 
is also required in the biosynthesis of 
phospholipids \cite{Cronan96} and coenzyme A \cite{Jackowski96}.

\begin{figure}[t]
\centering
\epsfig{file=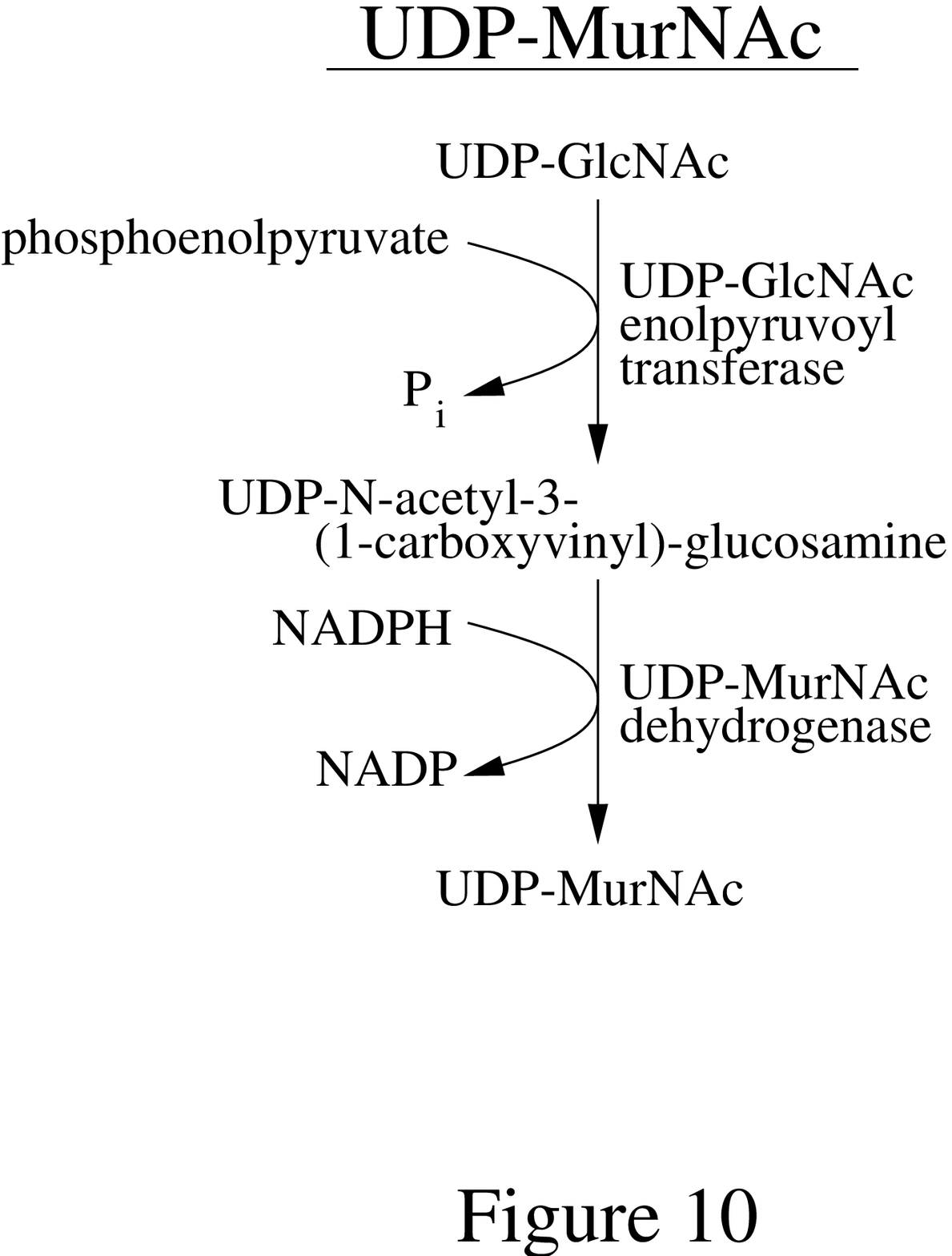,width=0.6\linewidth}
\caption{
Biosynthesis of UDP-MurNAc from UDP-GlcNAc. UDP-MurNAc
is required to initiate synthesis of peptidoglycan.}
\label{fig:mur}
\end{figure}

\begin{figure}[t]
\centering
\epsfig{file=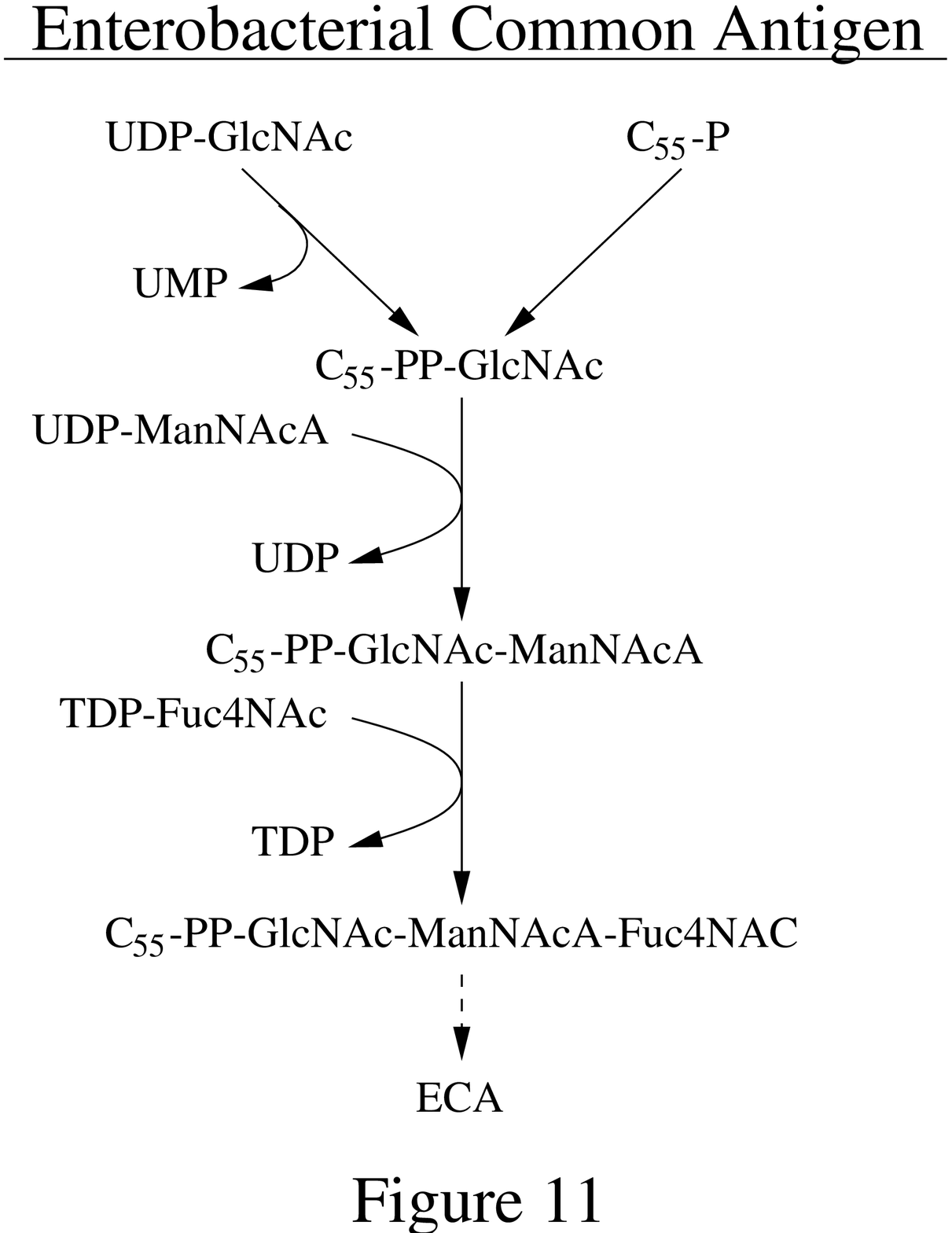,width=0.6\linewidth}
\caption{
Biosynthesis of enterobacterial common antigen (ECA). 
UDP-GlcNAc, a product of glutamine, is required
to inititiate ECA synthesis.}
\label{fig:eca}
\end{figure}

{\it Cell-surface polymers --} 
Similar to its role in nucleotide synthesis, glutamine is required for
synthesis of glucosamine from the common precursor D-fructose
6-P (Fig.~\ref{fig:pol}). Glucosamine 
is the substrate for synthesis of UDP-GlcNAc and
UDP-MurNAc (Fig.~\ref{fig:mur}). 
The latter is used to initiate synthesis of
the repeating unit of the osmotically-resistant layer of the cell
envelope, murein \cite{Park96}.  
UDP-GlcNAc is also used to initiate
synthesis of two of the three portions of the lipopolysaccharide (LPS)
component of the outer membrane, the lipid A and core portions
\cite{Raetz96}.
The biosynthesis of lipid A begins with the fatty acylation of
UDP-GlcNAc and later the core region is assembled 
using lipid A as a substrate.
In {\it E. coli}, UDP-GlcNAc is also required to initiate 
synthesis of the third portion of LPS, the O-antigen. 
UDP-galactose is used in place of UDP-GlcNAc in O-antigen
synthesis in {\it S. typhimurium}. Finally, as shown in 
Fig.~\ref{fig:eca}, UDP-GlcNAc is used to initiate 
synthesis of the trisaccharide repeating unit of the 
enterobacterial common antigen (ECA), a glycolipid
found in the outer membrane of enterobacteriaceae \cite{Meier90}.   
Depletion of glucosamine under nitrogen limitation would 
therefore potentially impact synthesis rates of the entire suite
of cell-surface polymers.

{\it Amino acids and proteins --}
In comparison to the above, 
the roles of glutamine in synthesis of the amino acids tryptophan,
histidine, arginine, and asparagine are less clear-cut, both in terms of the
positions of the glutamine-dependent reactions in their biosynthetic
pathways and in terms of the potential effects of depletion of these 
amino acids on growth.  Although glutamine is required to initiate
synthesis of tryptophan and is required for the synthesis of 
asparagine from aspartate, glutamine is utilized later in the arginine
and histidine biosynthetic pathways.  Given that ornithine, the co-substrate
of the glutamine-dependent (or more specifically carbamyl 
phosphate-dependent)
reaction in the arginine biosynthetic pathway (Fig.~9), is also a precursor of
polyamines, the position of the glutamine-dependent reaction can be
rationalized.  As noted in Results, this is not the case for the
histidine pathway, where glutamine is required at the fifth
dedicated step (Fig.~\ref{fig:his}).  
However, the enzyme involved in histidine
synthesis, imidazole glycerol-p synthase, has an
unusually low $K_m$ (0.24mM) for glutamine, suggesting that this
reaction is driven to completion even under nitrogen-limited 
conditions. 

Although depletion of the
tryptophan, histidine, arginine, and asparagine pools could certainly slow the
rate of elongation of proteins, it would not be expected to slow their
rate of initiation. Depletion of amino-acid pools therefore
runs the risk of stalling protein elongation in the midst of 
synthesis.  Interestingly, a requirement for glutamine by the
enzyme that catalyzes the first committed step in folate biosynthesis
might indirectly affect initiation of protein synthesis.  This enzyme,
aminodeoxychorismate synthase, has an unusually high $K_m$ (1.6mM)
for glutamine (Table 1) so that flux into this pathway will
be strongly reduced following glutamine depletion.  The co-factor, folate,
is required for one-carbon transfer reactions and hence is essential for
both the synthesis of methionine and for the formylation of methionine
that allows it to serve in the first step of protein synthesis.  Though
not in the context of nitrogen limitation, the possibility that
initiation of translation may be modulated by the folate pool 
has been noted by Gold \cite{Gold88}.

As outlined above, primary glutamine depletion is expected to
translate into secondary depletion of pools of one or more
glutamine products. Under conditions of depletion, the specific
regulatory systems for many of these product pathways 
({\it e.g.} tryptophan and histidine) 
are known to increase activity and synthesis of pathway
enzymes. Increased synthesis of these enzymes
due to depletion of product pools
would consume already scarce nitrogen without benefit and would
result in a futile tug-of-war between specific regulatory mechanisms and our
postulated general metabolic regulatory role for glutamine. Recent data
from DNA microarrays indicate that transcription of the tryptophan and
histidine biosynthetic operons is, in fact, slightly repressed under
nitrogen-limiting conditions, and hence that this tug-of-war does not
occur.  
For product pools that {\it rise} under nitrogen limitation,
repression of enzyme expression is expected. For the specific
product pools that deplete under nitrogen limitation, the regulatory
mechanisms may be designed to avoid upregulation of enzymes 
under this condition.
Interestingly, Rose and Yanofsky \cite{Yanofsky} 
described an uncharacterized
``override mechanism" that allows repression of the tryptophan operon
under conditions of slow growth, even when tryptophan pools are low.

To conclude, measurement of pools of glutamine products will shed light 
on how cellular homeostasis is maintained under nitrogen
limitation. Comparative measurements between pool sizes 
under maximal and slow growth conditions, rather than 
absolute measurements, will suffice for this purpose. 
Bioassays may therefore be appropriate to measure
relative pools of, {\it e.g.} folates \cite{Basso93}, 
and other products for which absolute assays 
are not practical.

We acknowledge valuable conversations with W.~van~Heeswijk,
J.~Ingraham, R.~A.~Lewis, R.~Simon, and D.~Zimmer.

\bigskip
\noindent Abbreviations used:\\
AICAR -- aminoimidazole carboxamide ribonucleotide,\\
C$_{55}$-P -- undecaprenyl monophosphate,\\
CoA -- coenzyme A,\\
ECA -- enterobacterial common antigen,\\
FGAM -- 5'-phosphoribosyl-N-formylglycinamidine,\\
FGAR -- 5'-phosphoribosyl-N-formylglycinamide,\\
Fuc4NAc -- 4-acetamido-4,6-dideoxy-D-galactose,\\
GlcNAc -- N-acetyl-D-glucosamine,\\
LPS -- lipopolysaccharide,\\
ManNAcA -- N-acetyl-D-mannosaminuronic acid,\\
MurNAc -- N-acetyl-D-muramic acid,\\
PRFAR -- phosphoribulosylformimino-AICAR-P,\\
PRPP -- phosphoribosyl pyrophosphate, \\
THF -- tetrahydrofolate,\\
THF(glu)$_n$ -- folylpolyglutamate,\\
XMP -- xanthosine monophosphate

\pagebreak

\newpage

\begin{table}
\footnotesize
\hspace*{-.5in}
\begin{tabular}{|llll|}\hline
\multicolumn{4}{|c|}{\bf Enzymes of Glutamine Metabolism}\\ \hline
\multicolumn{1}{|c|}{Enzyme}&
\multicolumn{1}{c|}{$K_M$ (mM)}&
\multicolumn{1}{c}{Other Substrate(s)}&
\multicolumn{1}{|c|}{Pathway(s)} \\ \hline

{\bf Amino acids:}&&&\  \\
Asparagine synthetase ({\it asnB}) & 0.66 \ \cite{Boehlein94} 
           & Aspartate, ATP & Asparagine \\
Glutamate synthase ({\it gltB,D}) & 0.4 \ \cite{Miller72} 
           & 2-oxoglutarate & Glutamate \\ 
Imidazole glycerol-P synthase ({\it hisF,H}) & 0.24 \  \cite{Klem93}
           & PRFAR & Histidine\\
Anthranilate synthase ({\it trpD,E}) & 0.5 \ \cite{Srinivasan73}  
           & Chorismate & Tryptophan \\
  &&&\  \\
{\bf Nucleotide bases:}& & & \ \\
Glutamine PRPP amidotransferase ({\it purF}) & 1.7 \ \cite{Zalkin93,Kim96}
           & PRPP & AMP, GMP \\
FGAM synthetase ({\it purL}) & 10.8\ \cite{Schendel89} 
           & FGAR, ATP & AMP, GMP \\
GMP synthetase ({\it guaA}) & 0.16 - 0.72 \ \cite{Zalkin77} 
           & XMP, ATP  & GMP \\ 
Carbamyl-P synthase ({\it carA,B}) & 0.4 \ \cite{Zalkin93,Meister89} 
           & HCO$_3^-$,  ATP(2) &  UTP, CTP, Arginine  \\
CTP synthetase ({\it pyrG}) & 0.16-1 \ \cite{Levitsky73} 
           &  UTP, ATP & CTP \\
  &&&\ \\

{\bf Amino sugar:}&\ &\ &\ \\
Glucosamine 6-P synthase ({\it glmS}) & 0.4 \  \cite{Badet87}
           & D-fructose 6-P & Glucosamine 6-P  \\
 &&& \ \\

{\bf Coenzyme:}&&&\ \\
Aminodeoxychorismate synthase ({\it pabA}) & 1.6 \ \cite{Viswanath95} 
           & Chorismate & Folic acid \\
\hline 

\end{tabular}

\normalsize

\caption{Enzymes of glutamine metabolism.  The $K_m$ for CTP
synthetase ranges from 0.16 mM with a saturating GTP concentration to
1 mM in the absence of GTP \protect\cite{Levitsky73}.
FGAM synthetase is inhibited by
glutamate according to $K_m \simeq 0.06 ( 1 + [{\rm glutamate}]/[1.6
{\rm mM}] )$ \protect\cite{Schendel89},
in the table we have taken [glutamate] = 20 mM.
GMP synthetase is sensitive to pH, with $K_m = 0.16\ {\rm mM}$ at pH =
7.5 and $K_m = 0.72\ {\rm mM}$ at pH = 8.2 \protect\cite{Zalkin77}.} 

\end{table}

\newpage

\ \\ 
\ \\
\ \\
\ \\
\ \\
\ \\
\begin{table}
\footnotesize
\hspace*{-.5in}
\begin{tabular}{|llll|}\hline
\multicolumn{4}{|c|}{\bf Enzymes of Glutamate Metabolism}\\ \hline
\multicolumn{1}{|c|}{Enzyme}&
\multicolumn{1}{c|}{$K_M$ (mM)}&
\multicolumn{1}{c}{Other Substrate(s)}&
\multicolumn{1}{|c|}{Pathway(s)} \\ \hline

{\bf Transaminase (TA) reactions:} &&&\ \\
Branched-chain-amino-acid TA ({\it ilvE}) 
	& &  2-keto-isovalerate  &  Alanine, Valine \\ 
	& &  2-keto-3-methyl-valerate  &  Iso-leucine \\
	& &  2-keto-4-methyl-pentanoate & Leucine \\
Acetylornithine TA ({\it argD}) 
	& &  N-acetylglutamyl-phosphate & Arginine \\
Aspartate TA ({\it aspC}) & 24 \cite{Kuramitsu90} 
	& Oxaloacetic acid & Aspartate \\
TDP-4-oxo-6-deoxy-D-glucose TA ({\it rffA}) & 5.1 \cite{Matsuhashi66} 
	& dTDP-4-dehydro-6-deoxy-D-glucose & TDP-fucose (ECA) \\
Histidinol-phosphate TA ({\it hisF,H}) &  10.4 \cite{Hsu89} 
	& Imidazole acetol-phosphate & Histidine \\
Aromatic-amino-acid TA ({\it tyrB}) & 0.28 \cite{Powell78}
	& Phenylpyruvate & Phenylalanine \\ 
	& & p-hydroxyphenylpyruvate & Tyrosine \\
Succinyldiaminopimelate TA ({\it dapC})& 5.2 \cite{Peterkofsky61}
	& N-succinyl-2-amino-6-ketopimelate
	& Lysine, Diaminopimelate \\
Phosphoserine TA ({\it serC}) 
	& & 3-phospho-hydroxypyruvate & Serine \\
& & & \\
{\bf Non-transaminase reactions:} & & & \\
N-acetylglutamate synthase ({\it argA}) & 5 \cite{Abdelal79} 
	& Acetyl-CoA & Arginine \\
Dihydrofolate synthase ({\it folC}) & 3.9 \cite{Bognar85} 
	& 7,8-dihydropteroate, ATP & Folic acid \\
Folylpolyglutamate synthase ({\it folC}) &  0.3 \cite{Bognar85} 
	& THF(glu)$_{n}$, ATP & Folic acid \\
Glutamine synthase ({\it glnA}) &5.5 \cite{Alibhai94} 
	& NH$_{3}$, ATP & Glutamine \\ 
Glutamate-cysteine lygase ({\it gshA}) & 0.5-1.7 \cite{Watanabe86,Huang88}
	& Cysteine, ATP & Glutathione \\ 
Glutamate racemase ({\it murI})& 4 \cite{Doublet94}& & Peptidoglycan \\ 
Glutamate 5-kinase ({\it proB})& 7-10 \cite{Baich69} 33 \cite{Smith84}
	& ATP & Proline  \\ \hline

\end{tabular}
\normalsize
\caption{Enzymes of glutamate metabolism. 
A putative transaminase acting directly from pyruvate to alanine has
not been purified \protect\cite{Reitzer96}.
The histidinol-phosphate transaminase is competitively inhibited
by imidazole acetol-phosphate; 
$K_m$ = 7 mM at non-inhibiting concentrations \protect\cite{Hsu89}.
N-acetylglutamate synthase $K_m = 5 mM$ in the absence of arginine, 
and increases with arginine concentration \protect\cite{Abdelal79}.}
\end{table}

\newpage

\begin{thebibliography}{99}

\bibitem{Zimmer00} Zimmer,~D.~P.,
Soupene,~E., Lee,~H.~L., Wendisch,~V.~F., Khodursky,~A.~B., Peter,~B.~J.,
Bender,~R.~A., \& Kustu,~S. (2000)         
{\it Proc. Nat. Acad. Sci.} {\bf 97}, 14674-14679.

\bibitem{Tempest70} Tempest,~D.~W., Meers,~J.~L., \& Brown,~C.~M. (1970)
{\it Biochem. J.} {\bf 117}, 405-407.

\bibitem{Reitzer90} Reitzer,~L.~J. (1990) in 
{\it Escherichia coli and Salmonella typhimurium}, ed. in chief,
Neidhardt,~F.~C. (ASM Press, Washington, D.~C.) Vol. 1, 
pp. 391-407.

\bibitem{Wohlhueter73} Wohlhueter,~R.~M., Schutt,~H., \& Holzer,~H.
(1973) {\it The Enzymes of Glutamine 
Metabolism}, eds. Prusiner,~S. \& Stadtman,~E.~R. (Academic Press, 
New York), pp. 45-64.

\bibitem{Ikeda96} Ikeda,~T.~P.,  Shauger,~A.~E., \&  Kustu,~S. (1996)
{\it J. Mol. Biol.} {\bf 259}, 589-607.

\bibitem{Neidhardt96}  
Neidhardt,~F.~C. (1996) ed. in chief, 
{\it Escherichia coli and Salmonella typhimurium}
(ASM Press, Washington, D.~C.).

\bibitem{Prusiner73}   Prusiner,~S. \& Stadtman,~E.~R. (1973)
eds. {\it The Enzymes of Glutamine 
Metabolism} (Academic Press, New York).

\bibitem{Raetz96} Raetz,~C.~R.~H. (1996) in 
{\it Escherichia coli and Salmonella typhimurium}, ed. in chief,
Neidhardt,~F.~C. (ASM Press, Washington, D.~C.) Vol. 1, 
pp. 1035-1063.


\bibitem{Reitzer96} Reitzer,~L.~J. (1996) in 
{\it Escherichia coli and Salmonella typhimurium}, ed. in chief,
Neidhardt,~F.~C. (ASM Press, Washington, D.~C.) Vol. 1, 
pp. 380-390. 

\bibitem{Zalkin93} Zalkin,~H. (1993) {\it Adv. Enzymol. Relat. Areas
Mol. Biol.} {\bf 66}, 203-309.

\bibitem{Stern99} Stern,~R.~J.,
Lee,~T.~Y., Lee,~T.~J., Yan,~W., Scherman,~M.~S., Vissa,~V.~D., 
Kim,~S.~K., Wanner,~B.~L.,  \& McNeil,~M.~R. (1999)
{\it Microbiol.} {\bf 145}, 663-671.

\bibitem{Davis92} Davis,~R.~H., Morris,~D.~R., \& Coffino,~P (1992)
{\it Microbiol. Rev.} {\bf 56}, 280-290.

\bibitem{Kern80} Kern,~D.,
Potier,~S., Lapointe,~J., \& Boulanger,~Y. (1980)
{\it Biochim. Biophys. Acta} {\bf 607}, 65-80.

\bibitem{Ibba96} Ibba,~M.,
Hong,~K.~W., Sherman,~J.~M., Sever,~S., \& Soll,~D. (1996)
{\it Proc. Nat. Acad. Sci.} {\bf 93}, 6953-6958.

\bibitem{Cronan96} Cronan,~J.~E.~Jr. and Rock,~C.~O. (1996) in 
{\it Escherichia coli and Salmonella typhimurium}, ed. in chief,
Neidhardt,~F.~C. (ASM Press, Washington, D.~C.) Vol. 1, 
pp. 612-636. 

\bibitem{Jackowski96} Jackowski,~S. (1996) in 
{\it Escherichia coli and Salmonella typhimurium}, ed. in chief,
Neidhardt,~F.~C. (ASM Press, Washington, D.~C.) Vol. 1, 
pp. 687-694. 

\bibitem{Park96} Park,~J.~T. (1996) in 
{\it Escherichia coli and Salmonella typhimurium}, ed. in chief,
Neidhardt,~F.~C. (ASM Press, Washington, D.~C.) Vol. 1, 
pp. 48-57.  

\bibitem{Meier90} Meier-Dieter,~U., 
Starman,~R., Barr,~K., Mayer,~H., \& Rick,~P.~D. (1990)
{\it J. Biol. Chem.} {\bf 265}, 13490-13497.





\bibitem{Gold88} Gold,~L. (1988) {\it Ann. Rev. Biochem.}
{\bf 57}, 199-233.

\bibitem{Yanofsky} Rose,~J.~K. \& Yanofsky,~C. (1972)
{\it J. Mol. Biol.} {\bf 69}, 103-18. 

\bibitem{Basso93} Basso,~J., Tiganos,~E., \& Herrington,~M.~B.
(1993) {\it Mol. Gen. Genet.} {\bf 238}, 218-224.


\bibitem{Boehlein94} Boehlein,~S.~K., Richards,~N.~G.~J.,
 \& Schuster,~S.~M. (1994) {\it J. Biol. Chem.} {\bf 269}, 7450-7457.

\bibitem{Miller72} Miller,~R.~E. \& Stadtman,~E.~R. (1972)
{\it J. Biol. Chem.} {\bf 247}, 7407-7419.

\bibitem{Klem93}  Klem,~T.~J. \&   Davisson,~V.~J. (1993) 
{\it Biochemistry} {\bf 32}, 5177-5186.

\bibitem{Srinivasan73} Srinivasan,~P.~R. (1973) in 
{\it The Enzymes of Glutamine 
Metabolism}, eds. Prusiner,~S. \& Stadtman,~E.~R. (Academic Press, 
New York) pp. 545-568.

\bibitem{Kim96}  Kim,~J.~H., Krahm,~J.~M., Tomchick,~D.~R.,
Smith,~J.~L., \& Zalkin,~H. (1996) {\it J. Biol. Chem.} {\bf 271}, 15549-15557.

\bibitem{Schendel89}  Schendel,~F.~J.,  Mueller,~E.,  Stubbe,~J.,
Shiau,~A.,  \&  Smith,~J.~M. (1989) {\it Biochemistry} {\bf 28}, 2459-2471.

\bibitem{Zalkin77}  Zalkin,~H. \&  Truitt,~C.~D. (1977) 
{\it J. Biol. Chem.} {\bf 252}, 5431-5436.

\bibitem{Meister89}  Meister,~A. (1989) 
{\it Adv. Enzymol. Relat. Areas Mol. Biol.} {\bf 62}, 315-74.

\bibitem{Levitsky73} Levitsky,~A. (1973) in {\it The Enzymes of Glutamine 
Metabolism}, eds. Prusiner,~S. \& Stadtman,~E.~R. (Academic Press, New York),
pp. 505-521.

\bibitem{Badet87} Badet,~B. Vermoote,~P.  Haumont,~P.-Y., Lederer,~F.,
\& Le Goffic,~F. (1987) {\it Biochemistry} {\bf 26}, 1940-1948.

\bibitem{Viswanath95} Viswanath,~V.~K.,  Green,~J.~M., \&  
Nichols,~B.~P. (1995) {\it J. Bacteriol.} {\bf 177} 5918-5923.

\bibitem{Kuramitsu90} Kuramitsu,~S., Hiromi,~K., Hayashi,~H.,
Morino,~Y., \& Kagamiyama,~H. (1990) 
{\it Biochemistry} {\bf 29}, 5469-5476.

\bibitem{Matsuhashi66} Matsuhashi,~M. \&
Strominger,~J.~L. (1966)
{\it J. Biol. Chem.} {\bf 241}, 4738-4744.

\bibitem{Hsu89} Hsu,~L.~C.,
Okamoto,~M., Snell,~E.~E. (1989)
{\it Biochimie} {\bf 71}, 477-489.   

\bibitem{Powell78} Powell,~J.~T. \&
Morrison,~J.~F. (1978)
{\it Eur. J. Biochem.} {\bf 87}, 391-400.   

\bibitem{Peterkofsky61} Peterkofsky,~B. \& 
C.~Gilvarg (1961)
{\it J. Biol. Chem.} {\bf 236}, 1432-1438.

\bibitem{Abdelal79} Abdelal,~A.~T.  \&
Nainan,~O.~V. (1979)
{\it J. Bacteriol.} {\bf 137}, 1040-1042.

\bibitem{Bognar85} Bognar,~A.~L.,
Osborne,~C., Shane,~B., Singer,~S.~C., \& Ferone,~R. (1985)
{\it J. Biol. Chem.} {\bf 260}, 5625-5630.

\bibitem{Alibhai94} Alibhai,~M. \& 
Villafranca,~J.~J. (1994)
{\it Biochemistry} {\bf 33}, 682-686.

\bibitem{Watanabe86} Watanabe,~K.,
Murata,~K., \& Kimura,~A. (1986)
{\it Agric. Biol. Chem.} {\bf 50}, 1925-1930.

\bibitem{Huang88} Huang,~C.~S.,
Moore,~W.~R, \& Meister,~A. (1988)
{\it Proc. Natl. Acad. Sci.} {\bf 85}, 2464-2468.

\bibitem{Doublet94} Doublet,~M.,
Heijenoort,~J.~v., \& Mengin-Lecreulx,~D. (1994)
{\it Biochemistry} {\bf 33}, 5285-5290.

\bibitem{Baich69} Baich,~A. (1969) {\it Biochim. Biophys. Acta}
    {\bf 192}, 462-467.

\bibitem{Smith84} Smith,~C.~J.,
Deutch,~A.~H, \& Rushlow,~K.~E. (1984)
{\it J. Bacteriol.} {\bf 157}, 545-551.

\end{thebibliography}
\end{document}